\documentclass{article}
\usepackage{amsfonts}
\usepackage{spconf,amsmath,graphicx}
\usepackage{amssymb}
\usepackage{epsfig}
\usepackage{cite}
\usepackage{color, soul}
\usepackage{array}
\usepackage{bm}
\usepackage{ifpdf}
\usepackage{amsthm}

\title{Performance of orthogonal matching pursuit for multiple measurement vectors}

\name{Jie Ding, Laming Chen, and Yuantao Gu$^*$
\thanks{This work was supported in part by the National Natural Science Foundation of
 China under Grants NSFC 60872087 and NSFC U0835003. The corresponding author of this paper is
 Yuantao Gu (gyt@tsinghua.edu.cn). }}
\address{State Key Laboratory on Microwave and Digital Communications\\
 Tsinghua National Laboratory for Information Science and Technology\\
 Department of Electronic Engineering, Tsinghua University, Beijing 100084, CHINA}
\begin{document}
\ninept

\maketitle

\begin{abstract}
In this paper, we consider orthogonal matching pursuit (OMP) algorithm
for multiple measurement vectors (MMV) problem.
The robustness of OMPMMV is studied under general perturbations---when the
measurement vectors as well as the sensing matrix are incorporated
with additive noise. The main result shows that although exact recovery of the sparse solutions is unrealistic in noisy scenario,
recovery of the support set of the solutions is
guaranteed under suitable conditions. Specifically, a sufficient condition is derived that guarantees
exact recovery of the sparse solutions in noiseless scenario.
\end{abstract}


\section{Introduction}
Finding the sparse solution to an under-determined linear equation is one of the most basic
problems in some fields of signal processing. It has also received significant attention since the
introduction of compressive sensing (CS), which is widely investigated in the past decade
\cite{Romberg}. The basic mathematical formula is
\begin{equation}\label{dj12}
\bm y=\bm \Phi \bm  x.
\end{equation}
In the field of CS, $\bm y \in \mathbb{C}^m$ denotes the measurement vector, $\bm \Phi\in
\mathbb{C}^{m\times n}$ with $m<n$ is called the sensing matrix, and $\bm x \in \mathbb{C}^n$ is
the sparse signal to be recovered, which means most of its entries are zero. (\ref{dj12}) is also
termed single measurement vector (SMV) problem. Many algorithms, including orthogonal matching pursuit (OMP), are proposed to
solve the SMV problem. The recovery performances of these algorithms are studied in several scenarios,
depending on whether the noise on $\bm y$ or $\bm \Phi$ exists. Many researches have been done on
the recovery accuracy of OMP in those scenarios \cite{Ding,Denis,Mark}.

In this paper, the recovery performance of computing sparse solutions to multiple measurement
vectors (MMV) is analyzed. The MMV problem was initially motivated by a neuromagnetic inverse
problem which is involved in Magnetoencephalography \cite{Rao}. In a recently proposed system named
the modulated wideband converter (MWC) \cite{Moshe}, MMV recovery also plays a important role in
detecting the locations of narrowband signals. The MMV problem is formulated as the following
equations
\begin{equation}\label{dj601}
\bm Y=\bm \Phi \bm X,
\end{equation}
where $\bm Y \in \mathbb{C}^{m\times L}$, $\bm \Phi \in \mathbb{C}^{m\times n}$, and $\bm X \in
\mathbb{C}^{n\times L}$. The matrix $\bm Y$ is made up of $L$ measurement vectors, and matrix $\bm
X$ is the sparse signal to be recovered. Here, a matrix $\bm X$ is called jointly $k$-sparse or
$k$-sparse if it contains no more than $k$ nonzero rows.

Recovery algorithms to the MMV problem include convex relaxation and OMPMMV, which is an extension
of OMP to the MMV \cite{J,Chen}. Other algorithms, including FOCal Underdetermined System Solver
(FOCUSS) \cite{Rao} and Joint $\ell_{2,0}$ Approximation Algorithm (JLZA) \cite{Hyder},
are also put forward. When $\bm Y$ and $\bm \Phi$ are unperturbed, the
recovery process of OMPMMV can be written as:
\begin{equation}\label{dj340}
(N_0)\qquad \tilde{\bm X}=R(\bm Y, \bm \Phi, k),
\end{equation}
where $R$ denotes the recovery process, $k$ denotes the sparsity level, and $\tilde{\bm X}$ is the
approximation of the original sparse signal $\bm X$. $(N_0)$ process is an ideal one. It has been
shown that under certain matrix-norm-based conditions, sparse signal can be exactly recovered
\cite{Chen}, i.e. $\tilde{\bm X}=\bm X$.

More generally, the perturbed observation matrix $\tilde{\bm Y}$ and sensing matrix $\tilde{\bm
\Phi}$ in the form of
\begin{equation}\label{dj413}
\tilde{\bm Y}=\bm Y+\bm B,\,\,\tilde{\bm \Phi}=\bm \Phi+\bm E.
\end{equation}
needs to be taken into account. In such scenario, the recovery process becomes
\begin{equation}\label{dj360}
(N_2)\qquad \tilde{\bm X}=R(\tilde{\bm Y}, \tilde{\bm \Phi}, k).
\end{equation}
Such consideration is necessary because perturbations on $\bm Y$ and $\bm \Phi$ always exist in
practice. In many CS scenarios, such as when $\bm \Phi$ represents a system model \cite{Moshe},
$\bm E$ denotes the system perturbation in realization. Also, $\bm B$ denotes the measurement
perturbation when quantization effects introduce considerable noise to $\bm Y$.

As far as we know, few researches have been done yet on the robustness of OMPMMV in $(N_2)$
process. In this paper, the corresponding theorem for SMV in \cite{Ding} is extended to MMV under
the general $(N_2)$ scenario. The main result shows that under certain conditions based on
the Restricted Isometry Property (RIP), the
locations of nonzero rows of $\bm X$ can be exactly recovered, and an upper bound on the recovery
error is given. In addition, it is demonstrated that such conditions guarantee exact recovery of
$\bm X$ in $(N_0)$ process.

\subsection{Notations and Assumptions}

Assume $\bm U \in \mathbb{C}^{n \times L}$ is a matrix, then its columns and rows are represented
as $\bm u_j,\,j=1,\ldots,L$ and $\bm U(i),\,i=1,\ldots ,n$, respectively. Notice that in this
paper, upper-case letters are used to denote matrices. The support set ${\rm supp}(\bm u)$ denotes
the indices of nonzero entries in vector $\bm u$. The support set of a matrix $\bm U$ is defined as
${\rm supp}(\bm U)=\bigcup_{j=1,\ldots,L}\textrm{supp}(\bm u_j)$. The number of elements in ${\rm
supp}(\bm u)$ and in ${\rm supp}(\bm U)$ are denoted as $\|\bm u\|_0$ and $\|\bm U\|_0$,
respectively.

The symbols $\sigma_{\max}(\cdot)$, $\sigma_{\min}(\cdot)$, $\|\cdot\|_2$, and
$\|\cdot\|_\textrm{F}$, denote the maximum, minimum nonzero singular values, spectral norm, and
Frobenius norm of a matrix, respectively. Let $\|\cdot\|_2^{(k)}$ denote the largest spectral norm
taken over all $k$-column submatrices. The perturbations $\bm E$ and $\bm B$ can be quantified with
the following relative bounds,
\begin{equation}\label{dj216}
\frac{\|\bm E\|_2}{\|\bm \Phi\|_2}\leq \varepsilon_0,\, \frac{\|\bm E\|_2^{(l)}}{\|\bm
\Phi\|_2^{(l)}}\leq \varepsilon,\, l=1,\ldots,k,\, \frac{\|\bm B\|_\textrm{F}}{\|\bm
Y\|_\textrm{F}}\leq \varepsilon_b.
\end{equation}
In this paper, it is assumed that $\varepsilon_0,\,\varepsilon$, and $\varepsilon_b$ are far less
than $1$.

\section{Background}

\subsection{Orthogonal Matching Pursuit for MMV (OMPMMV)}

The key idea of OMPMMV, which is similar to OMP, lies in the attempt to reconstruct the support set
$\Lambda$  of $\bm X$ iteratively by starting with $\Lambda=\emptyset$. The pseudo-code of OMPMMV
is described in Table~\ref{ompmmvalgorithm}. In fact, the process of OMPMMV can be mathematically
expressed as follows.

\begin{table}[t]
\renewcommand{\arraystretch}{1.1}
\caption{The OMPMMV Algorithm}
\begin{center}
\begin{tabular}{l}
\hline {\bf Input:}\label{ompmmvalgorithm}
\hspace{1em}$\bm Y$, $\bm \Phi$, $k$; \\
{\bf Initial state:} \hspace{0.5em}
$\bm R^0=\bm Y$, $\Lambda^0=\emptyset$, $l=0$;\\
{\bf Repeat}\\
\hspace{3.5em}$l=l+1$;\\
\hspace{1.5em}match step:\\
\hspace{3.5em}$\bm H^l=\bm \Phi^\textrm{T} \bm R^{l-1}$;\\
\hspace{1.5em}identify step:\\
\hspace{3.5em}$\Lambda^l=\Lambda^{l-1} \bigcup \{\textrm{arg}\max_j \|\bm H^l(j)\|_2 \}$;\\
\hspace{1.5em}update step:\\
\hspace{3.5em}$\bm X^l=\textrm{arg}\min_{\bm Z: \textrm{supp}(\bm
Z)\subseteq\Lambda^l}\|\bm Y- \bm \Phi \bm Z\|_\textrm{F}$;\\
\hspace{3.5em}$\bm R^l=\bm Y-\bm \Phi \bm X^l$;\\
{\bf Until} \hspace{1em} $l=k$;\\
{\bf Output:}
\hspace{0.5em}$\bm X^k$.\\
\hline
\end{tabular}
\end{center}
\end{table}

Suppose $\Lambda\subset\{1,\ldots,n\}$. Let $\bm u|_\Lambda$ denote the $|\Lambda|\times 1$ vector
containing the entries of $\bm u$ indexed by $\Lambda$. Let $\bm \Phi_\Lambda$ denote the $m \times
|\Lambda|$ matrix obtained by selecting the columns of sensing matrix $\bm \Phi$ indexed by
$\Lambda$. $\bm \Phi_\Lambda^\dag=(\bm \Phi_\Lambda^\textrm{T} \bm \Phi_\Lambda)^{-1} \bm
\Phi_\Lambda^\textrm{T}$ denotes the Moore-Penrose pseudoinverse of $\bm \Phi_\Lambda$. Define $\bm
P_\Lambda = \bm \Phi_\Lambda \bm \Phi_\Lambda^\dag$ and $\bm P_\Lambda^\bot = \bm I-\bm P_\Lambda$
as the orthogonal projection operator onto the column space of $\bm \Phi_\Lambda $ and its
orthogonal complement, respectively. Define $\bm A_\Lambda = \bm P_\Lambda^\bot \bm \Phi$. From the
theory of linear algebra, orthogonal projection operator $\bm P$ obeys $\bm P=\bm P^\textrm{T}=\bm
P^2$ and that the columns of $\bm A_\Lambda$ indexed by $\Lambda$ equal zero.

Now, suppose that OMPMMV performs at $l$th iteration, and $\Lambda^{l-1}$ is the estimation of
${\rm supp}(\bm X)$ from the previous iteration. The following discussion demonstrates the
generation of $\Lambda^l$.

In the update step of the previous iteration, which is actually solving a least squares problem, it
can be derived that
\begin{align}
\bm R^{l-1} &= \bm Y - \bm
\Phi_{\Lambda^{l-1}} \bm X^{l-1}|_{\Lambda^{l-1}}
= \bm Y - \bm \Phi_{\Lambda^{l-1}} \bm \Phi_{\Lambda^{l-1}}^\dag \bm Y \nonumber\\
&=(\bm I-\bm P_{\Lambda^{l-1}})\bm Y
=\bm P_{\Lambda^{l-1}}^\bot \bm \Phi \bm
X=\bm A_{\Lambda^{l-1}}\bm X.\label{dj214}
\end{align}
In the matching step, one has
\begin{align}\label{dj228}
\bm H^l = \bm \Phi^\textrm{T} (\bm
P_{\Lambda^{l-1}}^\bot)^\textrm{T} \bm P_{\Lambda^{l-1}}^\bot \bm Y
= \bm A_{\Lambda^{l-1}}^\textrm{T} \bm
A_{\Lambda^{l-1}} \bm X.
\end{align}
Then, in the identify step, $\Lambda^l=\Lambda^{l-1}\bigcup \{\textrm{arg}\max_j \|\bm
H^l(j)\|_2\}. $

From (\ref{dj214}), (\ref{dj228}), and the fact that the columns of $\bm A_\Lambda$ indexed by
$\Lambda$ equal zero, several useful conclusions can be derived.
\begin{enumerate}
\item
$\bm H^l(j)=\bm 0,\, \forall j \in\Lambda^{l-1}.$ Therefore $\textrm{arg}\max_j \|\bm H^l(j)\|_2
\notin \Lambda^{l-1}$, $|\Lambda^{l-1}|=l-1$.
\item
(\ref{dj228}) can be written as $\bm H^l=\bm A_{\Lambda^{l-1}}^\textrm{T} \bm A_{\Lambda^{l-1}} \bm
X^{*{l-1}}$, where $\bm X^{*{l-1}}|_{\Lambda^{l-1}}=\bm 0$ and $\bm
X^{*{l-1}}|_{(\Lambda^{l-1})^{\textrm{c}}}=\bm X|_{(\Lambda^{l-1})^{\textrm{c}}}$.
\end{enumerate}
It is easy to check that 1) still holds when $\bm Y$ and $\bm \Phi$ in the above analysis are
contaminated ones, i.e.
\begin{equation}\label{dj227}
\tilde{\bm H}^l(j)=\bm 0,\,\, \forall j \in\Lambda^{l-1}.
\end{equation}

\subsection{The Restricted Isometry Property  (RIP)}

For each integer $k=1,\ldots,n$, the RIP for any matrix $\bm A\in \mathbb{C}^{m\times n}$ defines
the restricted isometry constant (RIC) $\delta_k$ which is the smallest nonnegative number such
that
\begin{equation}
(1-\delta_k)\|\bm u\|_2^2 \leq \|\bm A \bm u\|_2^2 \leq (1+\delta_k)
\|\bm u\|_2^2
\end{equation}
holds for any $k$-sparse vector $\bm u$.
In other words, $\bm A$ acts as an approximate isometry on the set of $k$-sparse vectors for a small $\delta_k$.

\section{Contributions}

\textbf{Theorem 1:}  \quad Suppose that the inputs $\bm Y$ and $\bm
\Phi$ of OMPMMV algorithm are contaminated by noise as in (\ref{dj413}).
Define the relative perturbations $\varepsilon_0$, $\varepsilon$,
and $\varepsilon_b$ as in (\ref{dj216}).
Let $t_0 = \min_{j\in
\textrm{supp}(\bm X)} \,\|\bm X(j)\|_2 $ \,and
\begin{align}
\varepsilon_h =&
\frac{9(2+\varepsilon)\varepsilon}{12-8(1+\varepsilon)^2}
  \left( \|\bm \Phi\|_2^4 + \frac{2}{3}\|\bm \Phi\|_2^2 \right)  \|\bm \Phi\|_2 \|\bm Y\|_\textrm{F}\nonumber
  \\\label{dj33}
 &+ (\varepsilon_0+\varepsilon_b+\varepsilon_0 \varepsilon_b)\|\bm \Phi\|_2\|\bm Y\|_\textrm{F}.
\end{align}
If $\bm \Phi$ satisfies the RIP of order $k+1$ with isometry
constant
\begin{equation}\label{dj17}
\delta_{k+1} < Q(k,t_0/\varepsilon_h),
\end{equation}
then for any $k$-sparse signal $\bm X$, OMPMMV will recover the support
set of $\bm X$ exactly from $\tilde{\bm Y}$ and $\tilde{\bm \Phi}$
in $k$ iterations, and the error between $\bm X$ and the recovered
signal $\tilde{\bm X}$ can be bounded as
\begin{equation}\label{dj100}
\|\tilde{\bm X}-\bm X\|_\textrm{F} / \|\bm X\|_\textrm{F} < (\varepsilon+\varepsilon_b) F(1/\sqrt{k}).
\end{equation}
The functions in (\ref{dj17}) and (\ref{dj100}) are defined as follows:
\begin{align}
Q(u,v)=\frac{1}{2\sqrt{u}+1}-\frac{4\sqrt{u}}{2\sqrt{u}+1}\frac{1}{\left(2+\frac{\textstyle1}{\textstyle\sqrt{u}}\right)v-2},
\, u,v\in \mathbb{R}^{+},\nonumber
\end{align}
\begin{align}
F(w)=\sqrt{\frac{1+w}{2-(1+w)(1+\varepsilon)^2}},\,w\in
(0,1).\nonumber
\end{align}

\textit{Remark 1:}  The above conclusion generalizes the result of Theorem 1 in \cite{Ding}.
If $L=1$, then the observation matrix and the signal matrix are reduced to vectors, which becomes an
SMV problem.  Thus Theorem 1 here can be regarded as an extension from SMV to MMV framework.
Notice that the requirements in the above Theorem 1 and the correspondent theorem in \cite{Ding}
share many similarities in form. The differences include the definition
of $\varepsilon_b$, $t_0$, and the change from $\ell_2$ norm of $\bm y$ to Frobenius norm of $\bm Y$.
Besides, the estimation error of $\bm X$ is derived in a relative form.

\textit{Remark 2:} For OMPMMV in the perturbed scenario (\ref{dj413}),
it is unrealistic to achieve the exact recovery of $\bm X$, because the least square problem is actually solved with perturbed data.
However, one can still recover the support of $\bm X$. This is of great
significance in many practical applications, e.g. the MWC \cite{Moshe} mentioned before.

\textit{Remark 3:} An MMV problem can be considered as how to achieve sparse representations for SMVs simultaneously. However, by looking at $\varepsilon_b$, we show that MMV has an advantage over a simple combination of SMVs.
In (\ref{dj216}), $\varepsilon_b$ is defined as an upper bound on:
\begin{equation}\label{n3}
\|\bm B\|_\textrm{F} / \|\bm Y\|_\textrm{F} = \sqrt{ \sum_{j=1}^L \|\bm b_j\|^2_2 / \sum_{j=1}^L \|\bm y_j\|^2_2}.
\end{equation}
From SMV view, $\max_{j=1,\cdots,L} \{\|\bm b_j\|_2 / \|\bm y_j\|_2\}$ cannot be too large for support recovery. Under the MMV scenario, however, a very large $\|\bm b_j\|_2 / \|\bm y_j\|_2$ may be balanced by other $\|\bm b_i\|_2 / \|\bm y_i\|_2,\,i\neq j$, and therefore the support recovery is not influenced.

Though a completely perturbed
situation is considered in Theorem 1, it is
helpful to consider three specific situations. The following three corollaries are derived from Theorem 1.

\textbf{Corollary 1:}  \quad  Suppose that $\bm \Phi$ satisfies the RIP of order $k+1$ with isometry constant
\begin{equation}\label{dj500}
\delta_{k+1}<\frac{1}{2\sqrt{k}+1},
\end{equation}
then OMPMMV will recover $\bm X$ exactly from
$\bm Y$ and $\bm \Phi$ in $k$ iterations.

The proof of Corollary 1 is directly derived by setting $\bm B=0,\,\bm E=0$, and
$\varepsilon_0=\varepsilon=\varepsilon_b=0$. Thus $\varepsilon_h=0$, and (\ref{dj17}) reduces to (\ref{dj500}).

Because SMV is a special case of
MMV when $L=1$ holds, Corollary 1 generalizes the results of Davenport
\cite{Mark} and Liu \cite{Entao}.
It was shown in the SMV case that
$\delta_{k+1}<1/(3\sqrt{k})$ is sufficient for OMP to exactly
recover any $k$-sparse signal \cite{Mark} (Th.3.1). Later, Liu and
Temlyakov relaxed the bound on the isometry constant to
$1/((1+\sqrt{2})\sqrt{k})$ \cite{Entao} (Th.5.2).

\textbf{Corollary 2:}  \quad  Suppose that $\tilde{\bm Y}$, $\bm
\tilde{\bm \Phi}$, $t_0$, and $\varepsilon_b$ meet the assumptions
made in Theorem 1, and $\tilde{\bm \Phi}=\bm \Phi$, which means that
only the observation matrix is perturbed.  Define
\begin{equation}\label{dj421}
\varepsilon_{h_1} = \varepsilon_b \|\bm \Phi\|_2\|\bm Y\|_\textrm{F}.
\end{equation}
 If $\bm \Phi$ satisfies the RIP of order $k+1$ with isometry constant
\begin{equation}\label{dj419}
\delta_{k+1}<Q(k,t_0/\varepsilon_{h_1}),
\end{equation}
then OMPMMV will recover the support set of $\bm X$ exactly from
$\tilde{\bm Y}$ and $\bm \Phi$ in $k$ iterations.
The recovery error can be bounded as
\begin{equation}
\|\tilde{\bm X}-\bm X\|_\textrm{F}/\|\bm X\|_\textrm{F} < \varepsilon_b (\sqrt{k}+1)/\sqrt{k-1}.
\end{equation}

\textbf{Corollary 3:}  \quad  Suppose that $\tilde{\bm Y}$,
$\tilde{\bm \Phi}$, $t_0$, $\varepsilon_0$, and $\varepsilon$ meet
the assumptions made in Theorem 1, and $\tilde{\bm Y}=\bm Y$, which
means that only the sensing matrix is perturbed.  Define
\begin{align}
\varepsilon_{h_2} =&
\frac{9(2+\varepsilon)\varepsilon}{12-8(1+\varepsilon)^2}
  \left( \|\bm \Phi\|_2^4 + \frac{2}{3}\|\bm \Phi\|_2^2 \right)  \|\bm \Phi\|_2 \|\bm
  Y\|_\textrm{F}\nonumber\\\label{dj422}
 &+ \varepsilon_0 \|\bm \Phi\|_2\|\bm Y\|_\textrm{F}.
\end{align}
If $\bm \Phi$ satisfies the RIP of order $k+1$ with isometry
constant
\begin{equation}   \label{dj420}
\delta_{k+1}<Q(k,t_0/\varepsilon_{h_2}),
\end{equation}
then OMPMMV will recover the support set of $\bm X$ exactly from $\bm
Y$ and $\tilde{\bm \Phi}$ in $k$ iterations.
The recovery error can be bounded as
\begin{equation}
\|\tilde{\bm X}-\bm X\|_\textrm{F}/\|\bm X\|_\textrm{F} < \varepsilon F(1/\sqrt{k}).
\end{equation}

The above two corollaries suggest that the relative recovery error scales almost linearly with
the noise level $\varepsilon_b$ or $\varepsilon$ when $k$ is fixed.

\section{Proofs}

\subsection{Some Lemmas}

Before proceeding to the proof of Theorem 1, some helpful lemmas are provided first.

\textbf{Lemma 1:}  \quad Suppose vector $\textbf{a}, \textbf{b}, \textbf{c}\in \mathbb{R}^L$, then
$\|\textbf{a+b+c}\|_2\leq \|\textbf{a}\|_2+\|\textbf{b}\|_2+\|\textbf{c}\|_2$, i.e. for arbitrary
$a_i, b_i, c_i\in \mathbb{R},\quad i=1,\ldots,L$,
\begin{align*}
\quad&\sqrt{\sum_{j=1}^L (a_j+b_j+c_j)^2} \leq \sqrt{\sum_{j=1}^L a_j^2}+\sqrt{\sum_{j=1}^L
b_j^2}+\sqrt{\sum_{j=1}^L c_j^2}.
\end{align*}

\textbf{Lemma 2 (\cite{Mark}, Lemma 3.1 ):}  \quad Let $\bm u, \bm v\in\mathbb{R}^{n}$ be given,
and suppose that matrix $\bm \Phi$ satisfies the RIP of order $\max(\|\bm u-\bm v\|_0,\|\bm u+\bm
v\|_0)$ with isometry constant $\delta$. Then
$$
|\langle \bm \Phi \bm u,\bm \Phi \bm v \rangle -\langle \bm u,\bm v\rangle|\leq \delta\|\bm
u\|_2\|\bm v\|_2.$$

\textbf{Lemma 3 (\cite{Mark}, Lemma 3.2 ):}  \quad Suppose that $\bm \Phi$ satisfies the RIP of
order $k$ with isometry constant $\delta$, and let $\Lambda\subset \{1,\ldots,n\}$. If
$|\Lambda|<k$, then
$$
\left(1-\frac{\delta}{1-\delta}\right)\|\bm u\|^2_2\leq \|A_\Lambda
\bm u\|_2^2 \leq (1+\delta)\|\bm u\|_2^2
$$
holds for all $\bm u\in \mathbb{R}^n$ such that $\|\bm u\|_0\leq k-|\Lambda|$ and ${\rm supp}(\bm
u)\cap\Lambda=\emptyset$.

\textbf{Lemma 4:}  \quad Let $\Lambda \subset\{1,\ldots,n\}$ and suppose $\bm X^*\in
\mathbb{R}^{n\times L}$ with ${\rm supp}(\bm X^*)\cap\Lambda=\emptyset$. Define $\bm H=\bm
A_\Lambda^\textrm{T}  \bm A_\Lambda \bm X^*$. Then if $\bm \Phi$ satisfies the RIP of order $\|\bm
X^*\|_0+|\Lambda|+1$ with isometry constant $\delta$, it holds that
\begin{equation}
\|\bm H(j) - \bm X^*(j)\|_2 \leq \frac{\delta}{1-\delta} \|\bm X^*\|_{\rm F}
\end{equation}
for all $j \notin \Lambda $.

\begin{proof}
According to the definition of $\bm H$, it can be implied that
\begin{align}
\bm H(j)&=(\bm h_1(j),\ldots,\bm h_L(j))\nonumber \\
&=(\langle \bm A_\Lambda \bm x_1^*,\bm A_\Lambda \bm e_j
\rangle, \cdots,\langle
\bm A_\Lambda \bm x_L^*,\bm A_\Lambda \bm e_j \rangle),\label{dj1101}
\end{align}
where $\bm e_j$ denotes the $j$-th vector from the cardinal basis. Lemma~3 implies that the
restriction of $\bm A_\Lambda$ to the columns indexed by $\Lambda^c$ satisfies the RIP of order
$(\|\bm X^*\|_0+|\Lambda|+1)-|\Lambda|=\|\bm X^*\|_0+1$ with isometry constant $\delta/(1-\delta)$.

Consider the scenario of $j\notin \Lambda$. Because for all $i\in \{1,\ldots,L\}$, $\|\bm
x_i^*\pm \bm e_j\|_0\leq \|\bm x_i^*\|_0+1\leq\|\bm X^*\|_0+1$ and ${\rm supp}(\bm x_i^*\pm \bm
e_j) \cap \Lambda=\emptyset$, it can be concluded from Lemma~2 that
\begin{align}
|\bm h_i(j)-\bm x_i^*(j)|&=|\langle \bm A_\Lambda \bm x_i^*,\bm A_\Lambda \bm e_j
\rangle-\langle \bm x_i^*, \bm e_j\rangle| \nonumber\\
& \leq \frac{\delta}{1-\delta}\|\bm x_i^*\|_2\|\bm e_j\|_2 \nonumber \\
&=\frac{\delta}{1-\delta}\|\bm x_i^*\|_2,
\quad i=1,\ldots,L.\label{dj102}
\end{align}
Combining (\ref{dj102}) with (\ref{dj1101}), it can be easily derived that
\begin{align*}
\|\bm H(j) - \bm X^*(j)\|_2 &= \|(\bm h_1(j)-\bm x_1^*(j),\ldots,\bm h_L(j)-\bm x_L^*(j))\|_2\\
& \leq  \frac{\delta}{1-\delta}\|(\|\bm x_1^*\|_2,\ldots,\|\bm x_L^*\|_2)\|_2 \\
&= \frac{\delta}{1-\delta}\|\bm X^*\|_{\rm F}.
\end{align*}
\end{proof}

\textbf{Lemma 5:}  \quad Suppose that $\Lambda, \bm \Phi, \bm X^*, \bm H, \delta$ meet the
assumptions specified in Lemma 4, $C$ is a constant, and $\tilde{\bm H}$ is the matrix in the
matching step of OMPMMV which satisfies $\|\tilde{\bm H}-\bm H\|_{\rm F} \leq C$. If
\begin{equation}\label{dj104}
\max \limits_{j \in \{1,\ldots,n\}}\|\bm X^*(j)\|_2 > \frac{2\delta}{1-\delta} \| \bm X^* \|_{\rm
F}+2C,
\end{equation}
it is guaranteed that ${\rm arg}\max_j \|\tilde{\bm H}(j)\|_2 \in$ ${\rm supp}(\bm X^*)$.

\begin{proof}
According to Lemma~4, for all $j \notin \Lambda $, $\|\bm H(j) - \bm X^*(j)\|_2 \leq \delta \|\bm
X^*\|_{\rm F}/(1-\delta)$, thus
\begin{align}\label{dj103}
\|\tilde{\bm H}(j)-\bm X^*(j)\|_2 &\leq \|\tilde{\bm H}(j)- \bm H(j)\|_2 + \|\bm H(j) -
\bm X^*(j)\|_2 \nonumber\\
&\leq \|\tilde{\bm H}- \bm H\|_F + \|\bm H(j) - \bm X^*(j)\|_2 \nonumber\\
&\leq \frac{\delta}{1-\delta}\|\bm X^*\|_{\rm F} +C.
\end{align}
Then for indices $j \notin $ ${\rm supp}(\bm X^*)$, we will have $\|\tilde{\bm H}(j)\| _2 \leq
\delta\|\bm X^*\|_{\rm F}/(1-\delta) + C$ (Recall that $\tilde {\bm H}(j)=\bm 0 $ for $j \in
\Lambda $). If (\ref{dj104}) is satisfied, then there exists some $j \in {\rm supp}(\bm X^*)$ with
$\|\bm X^* (j)\|_2
> 2\delta \| \bm X^* \|_{\rm F}/(1-\delta)+2C$. From (\ref{dj103}) and the triangle inequality, it
can be concluded that for this index $j$, $\|\tilde{\bm H}(j)\|_2 > \delta\|\bm X^*\|_{\rm
F}/(1-\delta) + C$.
\end{proof}

\textbf{Lemma 6:} \quad Suppose that $\|\tilde{\bm H}^l-\bm H^l \|_{\rm F} \leq C$ holds in the
$l$-th ($l=1,\ldots,k$) iteration during OMPMMV running, and $\bm \Phi$ satisfies the RIP of order
$k+1$ with isometry constant
\begin{equation}\label{dj2}
 \delta_{k+1}<Q(k,t_0/C).
\end{equation}
Here, $C$ and $t_0 = \min_{j\in \textrm{supp}(\bm
X)} \,\|\bm X(j)\|_2 $ are two constants. Then OMPMMV will recover the support set of $\bm X$ exactly
from $\tilde{\bm Y}$ and $\tilde{\bm \Phi}$ in $k$ iterations.

\begin{proof}
The proof works by induction. To begin with, consider the first iteration where
$\Lambda^0=\emptyset$. Because $t_0\leq \|\bm X\|_{\rm F}$, one has
$$
\delta_{\|\bm X\|_0+1} \leq \delta_{k+1}<Q(k,t_0/C) \leq Q(k,\|\bm X\|_{\rm F}/C),
$$
and this implies that
$$
\frac{\|\bm X\|_{\rm F}}{\sqrt{k}} > \frac{2 \delta_{\|\bm X\|_0+1}}{1-\delta_{\|\bm X\|_0+1}}\|\bm
X\|_{\rm F} + 2C.
$$
Notice that $\max_{j \in \{1,\ldots,n\}}\|\bm X^*(j)\|_2$ is no less than $ \|\bm X\|_{\rm
F}/\sqrt{k}$. Therefore (\ref{dj104}) can be satisfied. Consequently, according to Lemma~5, it is
proved that
$$
{\rm arg}\max_j \|\tilde{\bm H}^1(j)\|_2 \in\, {\rm supp}(\bm X).
$$

Furthermore, consider the general induction step. Suppose that OMPMMV is at $l$th iteration and all
previous iterations succeed, which means that $\Lambda^{l-1}$ is a subset of ${\rm supp}(\bm X)$.
Recall that the intersection of ${\rm supp}(\bm X^{*{l-1}})$ and $\Lambda^{l-1}$ is empty,
$|\Lambda^{l-1}|$ equals $l-1$, and $\|\bm X^{*{l-1}}\|_0$ is not greater than $k-l+1$. Assume that
$\bm \Phi$ satisfies the RIP of order $\|\bm X^{*{l-1}}\|_0+|\Lambda^{l-1}|+1$ with isometry
constant $\delta^{l-1}$. Because
$$
\|\bm X^{*{l-1}}\|_0+|\Lambda^{l-1}|+1 \leq k+1,
$$
and from the fact that $t_0$ is not greater than $\|\bm X^{*{l-1}}\|_{\rm F}$, one has
\begin{equation}\label{dj35}
\delta^{l-1} \leq \delta_{k+1}<Q(k,t_0/C)\leq Q(k,\|\bm X^{*{l-1}}\|_{\rm F}/C),
\end{equation}
which implies that
$$
\frac{\|\bm X^{*{l-1}}\|_{\rm F}}{\sqrt{k}} >  \frac{2 \delta^{l-1}}{1-\delta^{l-1}}\|\bm
X^{*{l-1}}\|_{\rm F} + 2C.
$$
Notice that
\begin{equation}\label{dj37}
\max \limits_{j \in \{1,\ldots,n\}}\|\bm X^*(j)\|_2 \geq \frac{\|\bm X^{*{l-1}}\|_{\rm
F}}{\sqrt{k-l+1}} \geq \frac{\|\bm X^{*{l-1}}\|_{\rm F}}{\sqrt{k}}.
\end{equation}
Thus (\ref{dj104}) can be satisfied. Therefore, according to Lemma~5, it can be concluded that
$$
{\rm arg}\,\max_j \|\tilde{\bm H}^l(j)\|_2\in\, {\rm supp} (\bm X^{*{l-1}}),
$$
and thus
$$
\Lambda^l \subset {\rm supp}(\bm X),
$$
which completes the proof of induction.
\end{proof}

\subsection{Proof of Theorem 1}

\begin{proof}
Similar to the scenario of OMP, here we need to give the upper bound
on $\|\tilde{\bm H}^l-\bm H^l\|_\textrm{F}$ (for all $l\in\{1,\ldots,k\}$), and
then replace the $C$ in Lemma 6 with this bound. $\Delta \bm H^l =\tilde{\bm H}^l-\bm H^l$ can be
written as $\Delta \bm H^l=(\Delta \bm h_1^l,\ldots,\Delta
\bm h_L^l),$ where
\begin{align}
\Delta \bm h_i^l = \tilde{\bm h_i}^l-\bm h_i^l
=& \bm E^\textrm{T}(\bm I-\tilde{\bm P}_{\Lambda^{l-1}})\bm y_i
- \bm \Phi^\textrm{T}(\tilde{\bm P}_{\Lambda^{l-1}}-\bm P_{\Lambda^{l-1}})\bm y_i \nonumber\\
&\,+
\tilde{\bm \Phi}^\textrm{T}(\bm I-\tilde{\bm P}_{\Lambda^{l-1}})(\tilde{\bm y_i}-\bm y_i),\,i=1,\ldots,L.\nonumber
\end{align}
From the proof of (38) in \cite{Ding}, there is
\begin{align}
\|\Delta \bm h_i^l\|_2 \leq & \|\bm E^\textrm{T}\|_2 \|\bm I-\tilde{\bm P}_{\Lambda^{l-1}}\|_2
\|\bm y_i\|_2 \nonumber\\
&\,+ \|\bm \Phi^\textrm{T}\|_2
\|\tilde{\bm P}_{\Lambda^{l-1}}-\bm P_{\Lambda^{l-1}}\|_2 \|\bm y_i\|_2 \nonumber\\
&\,+
\|\tilde{\bm \Phi}^\textrm{T}\|_2 \|\bm I-\tilde{\bm P}_{\Lambda^{l-1}}\|_2
\|\tilde{\bm y_i}-\bm y_i\|_2 \nonumber\\
\leq & \frac{9(2+\varepsilon) \varepsilon }{12-8(1+\varepsilon)^2}
\left( \|\bm \Phi\|_2^4 +
\frac{2}{3}\|\bm \Phi\|_2^2\right)\|\bm \Phi\|_2 \|\bm y_i\|_2 \nonumber\\
&\,+ \varepsilon_0 \|\bm \Phi\|_2 \|\bm y_i\|_2 \nonumber+
(\varepsilon_0+1)\|\bm \Phi\|_2\|\bm b_i\|_2 \nonumber
\end{align}
Combining this and Lemma 1, one gets
\begin{align}
\|\Delta \bm H^l\|_\textrm{F} =& \sqrt{\sum_{j=1}^L \|\Delta \bm h_j^l\|_2^2} \nonumber\\
\leq & \frac{9(2+\varepsilon) \varepsilon }{12-8(1+\varepsilon)^2}
\left( \|\bm \Phi\|_2^4 +
\frac{2}{3}\|\bm \Phi\|_2^2\right)\|\bm \Phi\|_2 \sqrt{\sum_{j=1}^L \|\bm y_j\|^2_2}\nonumber \\
&\,+\varepsilon_0 \|\bm \Phi\|_2 \sqrt{\sum_{j=1}^L \|\bm y_j\|^2} + (\varepsilon_0+1)\|\bm \Phi\|_2 \sqrt{\sum_{j=1}^L \|\bm b_j\|^2_2} \label{n1}.
\end{align}
Notice that
\begin{equation}\label{n2}
\sqrt{\sum_{j=1}^L \|\bm y_j\|^2_2} =\|\bm Y\|_\textrm{F},\,\sqrt{\sum_{j=1}^L \|\bm b_j\|^2_2} =\|\bm B\|_\textrm{F} \leq \varepsilon_b \|\bm Y\|_\textrm{F} .
\end{equation}
Applying (\ref{n2}) to (\ref{n1}), one gets the upper
bound on $\|\Delta \bm H^l\|_\textrm{F}$:
\begin{align}
\|\Delta \bm H^l\|_\textrm{F}
\leq & \frac{9(2+\varepsilon) \varepsilon }{12-8(1+\varepsilon)^2} \left(\|\bm \Phi\|_2^4
+ \frac{2}{3}\|\bm \Phi\|_2^2\right)\|\bm \Phi\|_2 \|\bm Y\|_\textrm{F} \nonumber\\
&\,+ (\varepsilon_0 + \varepsilon_b + \varepsilon_0 \varepsilon_b) \|\bm \Phi\|_2  \|\bm Y\|_\textrm{F}. \nonumber
\end{align}

At the end of the proof, (\ref{dj100}) is proved as follows.
According to (\ref{dj601}) and (\ref{dj413}), there is
$$\tilde{\bm Y}=\tilde{\bm \Phi} \bm X - \bm E \bm X + \bm B.$$
Because $\Lambda={\rm supp}(\bm X)$ is exactly recovered, one has
$$
\tilde{\bm X}|_{\Lambda} = \tilde {\bm \Phi}_\Lambda^\dag \tilde{\bm
Y} =\bm X|_{\Lambda} + \tilde {\bm \Phi}_\Lambda^\dag (- \bm E \bm X +
\bm B).
$$
Thus
\begin{align}
\frac{\|\tilde{\bm X} - \bm X \|_\textrm{F}}{\|\bm X\|_\textrm{F}} &\leq \|\tilde {\bm
\Phi}_\Lambda^\dag\|_2 \frac{\|\bm E\|^{(k)}_2 \|\bm X\|_\textrm{F} + \|\bm
B\|_\textrm{F}}{\|\bm X\|_\textrm{F}}\nonumber\\
&\leq \frac{1}{\sqrt{1-\tilde{\delta_k}}}\left(\|\bm E\|^{(k)}_2
+\varepsilon_b
\frac{\|\bm Y\|_\textrm{F}}{\|\bm X\|_\textrm{F}} \right)\nonumber\\
&\leq \frac{\|\bm E\|^{(k)}_2 +\varepsilon_b \|\bm Y\|_\textrm{F} / \|\bm
X\|_\textrm{F}}{\sqrt{2-(1+\delta_k)(1+\varepsilon)^2}}.\nonumber
\end{align}
Notice that
$$
\frac{\|\bm Y\|_\textrm{F}}{\|\bm X\|_\textrm{F}}=\frac{\|\bm \Phi \bm X \|_\textrm{F}}{\|\bm
X\|_\textrm{F}} \leq \frac{\|\bm \Phi\|_2^{(k)} \|\bm X \|_\textrm{F}}{\|\bm
X\|_\textrm{F}} \leq \sqrt{1+\delta_k},
$$
it can be concluded that
\begin{equation}\label{dj101}
\frac{\|\tilde{\bm X} - \bm X \|_\textrm{F}}{\|\bm X\|_\textrm{F}} \leq
(\varepsilon+\varepsilon_b) \sqrt{\frac{1+\delta_k}{2-(1+\delta_k)(1+\varepsilon)^2}}=(\varepsilon+\varepsilon_b) F(\delta_k).
\end{equation}
It has been mentioned in \cite{Mark} that the theoretical upper
bound of $\delta_{k+1}$ is $1/\sqrt{k}$ for exact recovery of
support set, if $\delta_{k+1}$ is used as a sufficient condition for
recovery of $\bm x$. Thus, $\delta_k \leq \delta_{k+1} <
1/\sqrt{k}$. Applying this to (\ref{dj101}), one finally gets
(\ref{dj100}).

\end{proof}

\section{Conclusion}
In this paper, robustness of OMPMMV under
general perturbations, which are in the
form of $\tilde{\bm Y}=\bm Y+\bm B$ and $\tilde{\bm \Phi}=\bm
\Phi+\bm E$, is studied. Though exact recovery of the sparse solutions
$\bm X$ from $\tilde{\bm Y}$ and $\tilde{\bm \Phi}$ is no longer realistic, Theorem
1 shows that exact recovery of the support set can be guaranteed under suitable conditions, which is important
in many practical applications. Furthermore, the
recovery error is bounded. This completely perturbed framework
extends some prior work in \cite{Mark,Entao,Ding} from SMV to MMV.


\begin{thebibliography}{1}                                   

\bibitem{Romberg}
E.~Cand\`{e}s, J.~Romberg, and T.~Tao, ``Robust uncertainty principles: Exact signal reconstruction
from highly incomplete frequency information,'' \emph{IEEE Trans. Information Theory}, vol.~52,
no.~2, pp.~489-509, Jan.~2006.

\bibitem{Denis}
L.~Denis, D.~A.~Lorenz, and D.~Trede, ``Greedy solution of ill-posed problems: Error bounds and
exact inversion,'' \emph{Inverse Problems}, vol.~25, no.~11, Nov.~2009.

\bibitem{Mark}
M.~A.~Davenport and M.~B.~Wakin, ``Analysis of orthogonal matching pursuit using the restricted
isometry property,'' \emph{IEEE Trans. Information Theory}, vol.~56, no.~9, pp.~4395-4401,
Sep.~2010.

\bibitem{Ding}
J.~Ding, L.~Chen, and Y.~Gu, ``Performance analysis of orthogonal matching pursuit under general
perturbations,'' submitted, available online:
http://arxiv.org/abs/1106.3373.

\bibitem{Rao}
S.~F.~Cotter, B.~D.~Rao, K.~Engan, and K.~Kreutz-Delgado, ``Sparse solutions to linear inverse
problems with multiple measurement vectors,'' \emph{IEEE Trans. Signal Processing}, vol.~53, no.~7,
pp.~2477-2488, Jul.~2005.

\bibitem{Moshe}
M.~Mishali and Y.~C.~Eldar, ``From theory to practice: Sub-Nyquist sampling of sparse wideband
analog signals,'' \emph{IEEE Journal of Selected Topics in Signal Processing}, vol.~4, no.~2,
pp.~375-391, Apr.~2010.


\bibitem{J}
J.~A.~Tropp, ``Algorithms for simultaneous sparse approximation. Part I: Greedy pursuit,''
\emph{Signal Process. (Special Issue on Sparse Approximations in Signal and Image Processing)},
vol.~86, pp.~572¨C588, Apr.~2006.


\bibitem{Chen}
J.~Chen and X.~Huo, ``Theoretical results on sparse representations of multiple-measurement
vectors,'' \emph{IEEE Trans. Signal Processing}, vol.~54, no.~12, pp.~4634-4643, Dec.~2006.

\bibitem{Hyder}
M.~M.~Hyder and K.~Mahata, ``A robust algorithm for joint-sparse recovery,'' \emph{IEEE Signal Processing Letters},
vol.~16, no.~12, pp.~1091-1094, Dec.~2009.

\bibitem{Entao}
E.~Liu and V.~N.~Temlyakov, ``Orthogonal super greedy algorithm and applications in compressed
sensing,'' preprint, 2010.


\end{thebibliography}
\end{document}